\documentclass[Conference,a4paper]{IEEEtran}
\IEEEoverridecommandlockouts
\usepackage{color}
\usepackage{graphicx}
\usepackage{epstopdf}
\usepackage{amsmath}
\usepackage{amssymb}
\usepackage{algorithm}
\usepackage{algorithmic}
\usepackage{amsmath}
\usepackage{multirow}
\usepackage{booktabs}
\usepackage{array}
\usepackage{amsthm}
\usepackage{stfloats}
\usepackage{caption}
\usepackage{subfigure}
\usepackage{bm}
\usepackage{booktabs}
\usepackage{setspace}
\usepackage{diagbox}
\usepackage{enumerate}
\usepackage{ulem}
\usepackage{geometry}
\allowdisplaybreaks[4]

\newenvironment{shrinkeq}[1]%
{\bgroup
 \addtolength\abovedisplayshortskip{#1}
 \addtolength\abovedisplayskip{#1}
 \addtolength\belowdisplayshortskip{#1}
 \addtolength\belowdisplayskip{#1}
 }
{\egroup\ignorespacesafterend}

\newcommand{\be}{\begin{equation}}
\newcommand{\ee}{\end{equation}}
\newcommand{\bea}{\begin{eqnarray}}
\newcommand{\eea}{\end{eqnarray}}
\newcommand{\ba}{\begin{array}}
\newcommand{\ea}{\end{array}}

\title{Joint Beamforming for RIS Aided Full-Duplex Integrated Sensing and Uplink Communication}
\author{\IEEEauthorblockN{Yuan Guo$^{1}$, Yang Liu$^{1\ast}
\thanks{ {$\ast$}\ Corresponding author.}
\thanks{The work of Yang Liu was supported in part by Grant DUT20RC(3)029 and in part by the Open Research Project Programme of the State Key Laboratory of Internet of Things for Smart City (University of Macau) under Grant SKLIoTSC(UM)-2021-2023/ORP/GA01/2022.}$,
Qingqing Wu$^{2}$,
Xin Zeng$^{3}$,
and
Qingjiang Shi$^{4}$
\vspace{-0.0 cm} }\\ 
\IEEEauthorblockA{$^{1}$School of Info. and Commun. Eng., Dalian University of Technology, Dalian, China  } \\
\IEEEauthorblockA{$^{2}$Department of Electronic Engineering, Shanghai Jiao Tong University, Shanghai, China} \\
\IEEEauthorblockA{$^{3}$College of Electronic and Information Engineering, Tongji University, Shanghai, China} \\
\IEEEauthorblockA{$^{4}$School of Software Eng., Tongji Univ. Shanghai, Shenzhen Research Institute of Big Data, Shenzhen, China} \\
 E-mails: {yuanguo@mail.dlut.edu.cn, yangliu\_613@dlut.edu.cn, {qingqingwu@sjtu.edu.cn},\\
 zengxin1@tongji.edu.cn,
 shiqj@tongji.edu.cn} }

\begin{document}
\maketitle
\pagestyle{empty}
\thispagestyle{empty}

\begin{abstract}
This paper studies integrated sensing and communication (ISAC) technology in a full-duplex (FD) uplink communication system.
As opposed to the half-duplex system,
where sensing is conducted in a first-emit-then-listen manner,
FD ISAC system emits and listens simultaneously and hence conducts uninterrupted target sensing.
Besides,
impressed by the recently emerging reconfigurable intelligent surface (RIS) technology,
we also employ RIS to improve the self-interference (SI) suppression and signal processing gain.
As will be seen,
the joint beamforming,
RIS configuration and mobile users' power allocation  is a difficult optimization problem.
To resolve this challenge,
via leveraging the cutting-the-edge majorization-minimization (MM) and penalty-dual-decomposition (PDD) methods,
we develop an iterative solution that optimizes all variables via using convex optimization techniques.
Numerical results demonstrate the effectiveness of our proposed solution and the great benefit of employing RIS in the FD ISAC system.

\end{abstract}


\maketitle
\section{Introduction}
Very recently, the integrated sensing and communication (ISAC) system has been cast with great attentions from both industry and academia \cite{ref1} $-$ \cite{ref2}.
The ISAC system aims at realizing both radar sensing and communication functionalities using one unified hardware set and overlapping spectrums.
The ISAC technology is highly meaningful in improving hardware and spectral efficiency and making wireless devices more amiable to diverse internet of things (IoT) applications in future.

\normalem
At the same time, the recently rising reconfigurable intelligent surface (RIS) \cite{ref3},
which is also widely known as intelligent reflecting surface (IRS) \cite{ref4},
has also been envisioned as a viable solution to enhance the next generation wireless communication system.
Via reflecting the electromagnetic waves and wisely adjusting their phase shifts, the RIS entitle the wireless network with a novel \emph{passive} beamforming capability at a low hardware and energy cost \cite{ref3}$-$\cite{ref4}.
The potentials and versatility of RIS has been extensively explored recently,
e.g.,
see \cite{ref3}$-$\cite{ref4} and the reference therein.

In light of the recent booming  RIS technology,
a number of latest works are exploring to employ RIS to enhance the ISAC system's performance
\cite{ref5}$-$\cite{ref10}.
For instance,
the authors in \cite{ref5} and  \cite{ref6}
studied minimizing multi-user interference (MUI)
while satisfying the beampattern and the Cram\'{e}r-Rao bound (CRB) constraints, respectively.
The work  \cite{ref7}
investigated the radar output signal-to-noise-ratio (SNR) maximization problem under the communication
quality of service (QoS) constraint.
The paper  \cite{ref9}
maximized  the weighted sum of the radar  SNR
and the  communication  SNR  in an RIS-aided ISAC system.
The recent literature  \cite{ref10}
investigated the transmit power minimization while assuring the cross-correlation pattern design.
Note that the existing literature usually considers  downlink communications and classical  half-duplex (HD) BS.

In this paper,
we consider an  RIS aided full-duplex (FD) ISAC system,
where multiple mobile users operate uplink  (UL) communication with the assistance of RIS employed in the vicinity.
We aim to maximize the mobile users' throughput
while assuring the radar sensing quality is maintained above a reasonable level.
The main contributions of this paper are elaborated as follows
\begin{itemize}
\item
This paper considers an RIS aided FD ISAC system,
where the UL communication
and mobile users' power allocation is considered,
which has rarely been discussed in relevant literature.

\item
The joint optimization of the active and passive beamforming and power allocation is highly challenging.
By virtue of majorization-minimization (MM) \cite{ref15} and penalty-dual-decomposition (PDD) \cite{ref16} methods,
we propose an alternative optimization solution that updates all variables efficiently.

\item
Numerical results show that our solution can effectively improve the system performance
and the deployment of RIS can significantly benefit the FD ISAC system.

\end{itemize}

\section{System Model and Problem Formulation}
\subsection{System Model}
\begin{figure}[t]
	\centering
	\includegraphics[width=.3\textwidth]{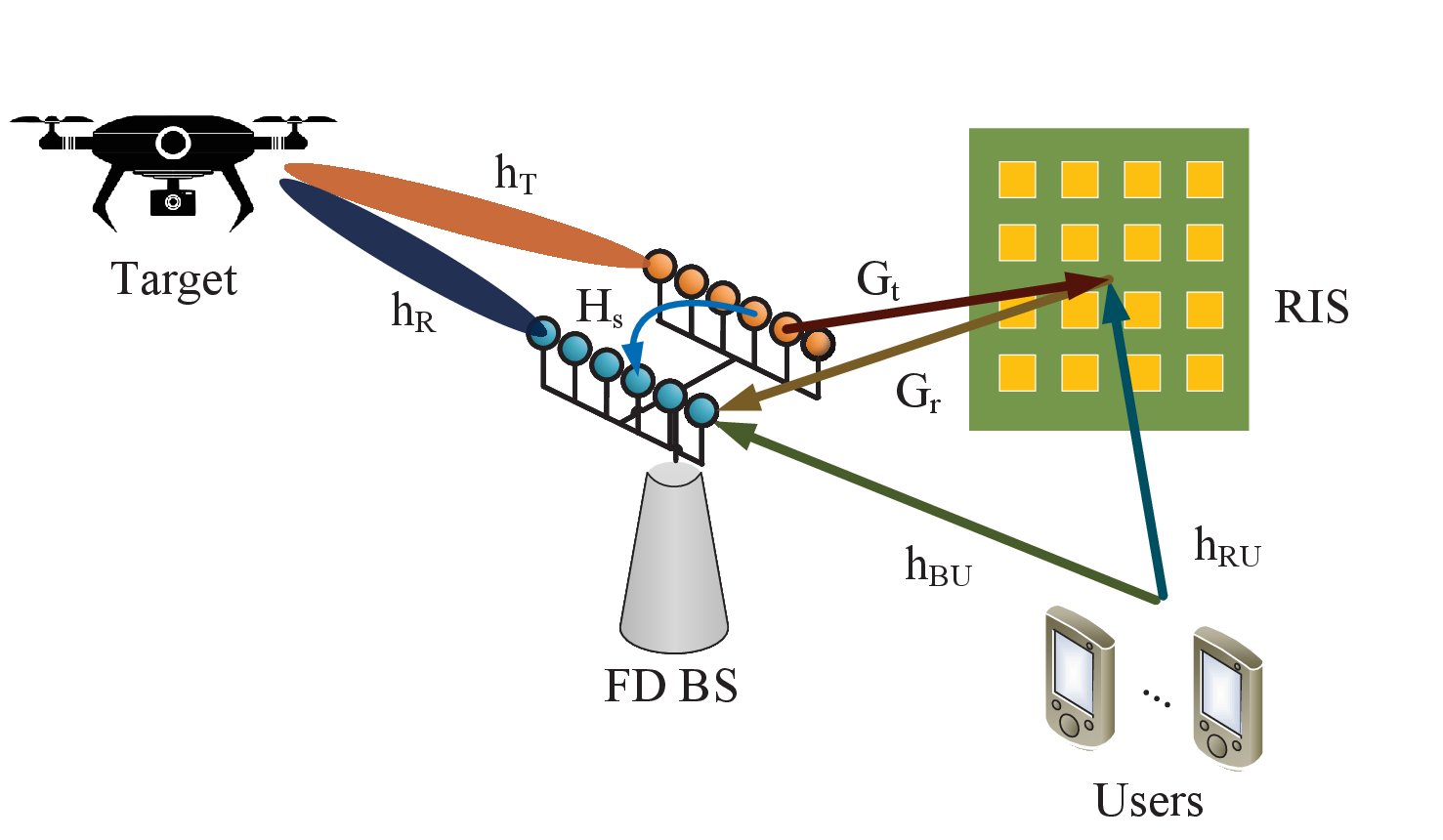}
	\caption{An  RIS-aided  FD ISAC  system.}
	\label{fig.1}
\end{figure}
As shown in Fig. 1,
an RIS aided FD ISAC network is considered,
where a FD BS equipped $N_t$ transmit (TX) antennas and $N_r$ receive (RX) antennas
serves $K$ single-antenna  mobile users with the help of an RIS consisting $M$ reflecting units and detects a point-like target simultaneously.
For convenience,
the sets of users and RIS units are denoted by $\mathcal{K}=\{1,\cdots,K\}$ and $\mathcal{M}=\{1,\cdots,M\}$, respectively.
RIS is employed in the vicinity of mobile users.

To conduct target sensing, the BS emits probing signal towards the target from its TX antennas and simultaneously receives the rebounding echoes from the target by its RX antennas. Meanwhile, the FD BS also executes communication functionality by receiving the UL signals transmitted from the mobile users.

Specifically,
the probing signal transmitted by the BS can be represented as
\begin{shrinkeq}{-0.5em}
\begin{align}
\mathbf{x}=\mathbf{W}\mathbf{s}_r,
\end{align}
\end{shrinkeq}
where
the vector
$\mathbf{s}_r \in \mathbb{C}^{N_t\times 1}$
is the probing signal and has zero mean and covariance matrix
$\mathbb{E}\{\mathbf{s}_r\mathbf{s}_r^H\}=\mathbf{I}_{N_t}$
and
the matrix $\mathbf{W}\in \mathbb{C}^{N_t\times N_t}$ denotes the digital baseband beamformer  for the probing signal.

The UL signal of $k$-th UL user is given as
\begin{shrinkeq}{-0.5em}
\begin{align}
x_{u,k}=\sqrt{q_k}s_{u,k}, \forall k \in \mathcal{K},
\end{align}
\end{shrinkeq}
where
$s_{u,k}$ and $q_k$ are the information symbol and transmission power of $k$-th user, respectively.
For simplicity,
we assume that $s_{u,k}$ are
circularly symmetric complex Gaussian (CSCG) random variables with zero mean and unit covariance,
i.e., $s_k\sim\mathcal{CN}(0,1)$.

At the same time,
the received signals at the BS is expressed as
\begin{align}
\mathbf{y}=&{\sum}_{k=1}^{K}(\mathbf{h}_{BU,k}+\mathbf{G}_r^H\bm{\Phi}\mathbf{h}_{RU,k})x_{u,k}\\ \nonumber
&+\alpha\mathbf{h}_{R}\mathbf{h}_{T}^H\mathbf{x}
+(\mathbf{G}_r^H\bm{\Phi}\mathbf{G}_t+\mathbf{H}_{s}^H)\mathbf{x}+\mathbf{n}_{BS},
\end{align}
where
$\mathbf{G}_t \in \mathbb{C}^{M \times N_t}$,
$\mathbf{G}_r \in \mathbb{C}^{M \times N_r}$,
$\mathbf{h}_{BU,k} \in \mathbb{C}^{N_r \times 1}$,
$\mathbf{h}_{RU,k} \in \mathbb{C}^{M \times 1}$,
$\mathbf{h}_T \in \mathbb{C}^{N_t \times 1}$,
$\mathbf{h}_R \in \mathbb{C}^{N_r \times 1}$
and
$\mathbf{H}_s \in \mathbb{C}^{N_t \times N_r}$
represent
the BS-RIS,
the RIS-BS,
the BS-$k$th  user,
the RIS-$k$th  user,
the BS-target,
the target-BS
and
the self-interference (SI)
channels,
respectively.
In our setting,
we assume that the RIS is employed in the vicinity of mobile users
while is far away from the radar target.
Besides,
signals experienced reflections more than two times are neglected due to significant attenuation.
The diagonal matrix
$\bm{\Phi}\triangleq\mathrm{Diag}(\bm{\phi})$  represents the RIS' reflection coefficients,
where the complex vector
${\bm{\phi}}=[e^{j\theta_1},\dots,e^{j\theta_M}]^T$ is
the phase-shifting conducted by the RIS elements to their impinging signals,
with $\theta_m$ representing the phase shift of $m$-th reflecting unit,
$\theta_m \in [0, 2\pi)$ and $\forall m \in \mathcal{M}$.
The coefficient $\alpha$  denotes the target radar cross section (RCS) and $\mathbb{E}\{\vert \alpha\vert^2\}= \sigma_{t}^2$,
and the CSCG random vector $\mathbf{n}_{BS}\sim \mathcal{CN}(0,\mathbf{I}_{N_r})$
represents the receiving noise at the BS.


To recover different users' information and improve target sensing performance,
the BS utilizes $K+1$ linear filter
$\mathbf{u}_j \in \mathbb{C}^{N_r \times 1}$,
$ \forall j \in \mathcal{{J}} \triangleq \{0\}\bigcup\mathcal{K}$
to post-process the received signal,
where index $0$ corresponds to the radar sensing filter bank.
Therefore,
the output of the $j$-th post-processor is given as
\begin{shrinkeq}{-0.2em}
\begin{align}
{y}_j = \mathbf{u}_j^H\mathbf{y},\ \forall j \in \mathcal{{J}}.
\end{align}
\end{shrinkeq}

Therefore,
the signal-to-interference-and-noise-ratio
(SINR) of the output of the radar and mobile users' post-processor are respectively given as
\begin{shrinkeq}{-0.2em}
\begin{small}
\begin{align}
&\textrm{SINR}_{r}(\mathbf{W},\mathbf{u}_0,\{q_k\},\bm{\phi})=\\
&\qquad\quad\quad\frac{\Vert\mathbf{u}^H_0\mathbf{H}\mathbf{W}\Vert^2_2
}
{\sum_{k=1}^{K} q_k\vert\mathbf{u}^H_0\mathbf{h}_{U,k}\vert^2
+\Vert\mathbf{u}^H_0\mathbf{G}\mathbf{W}\Vert^2_2
+\sigma^2_r\Vert\mathbf{u}_0^H\Vert^2_2},\nonumber\\
&\textrm{SINR}_{U,k}(\mathbf{W},\mathbf{u}_k,\{q_k\},\bm{\phi})=\\
&\frac{q_k\vert\mathbf{u}^H_k\mathbf{h}_{U,k}\vert^2}
{\sum_{i\neq k}^{K} q_i\vert\mathbf{u}^H_k\mathbf{h}_{U,i}\vert^2
+\Vert\mathbf{u}^H_k\mathbf{H}\mathbf{W}\Vert^2_2
+\Vert\mathbf{u}^H_k\mathbf{G}\mathbf{W}\Vert^2_2
+\sigma^2_r\Vert\mathbf{u}_k^H\Vert^2_2
},\nonumber
\end{align}
\end{small}
\end{shrinkeq}
\vspace{-0.2cm}

\noindent
where
$\mathbf{H}\triangleq\alpha\mathbf{h}_R\mathbf{h}_T^H$,
$\mathbf{G}\triangleq\mathbf{G}_r^H\boldsymbol{\Phi}\mathbf{G}_t+\mathbf{H}_{s}^H$
and
$\mathbf{h}_{U,k} \triangleq \mathbf{h}_{BU,k}+\mathbf{G}^H_r\boldsymbol{\Phi}\mathbf{h}_{RU,k}$.

The UL communication rate of the  $k$-th user is given by
\begin{shrinkeq}{-0.2em}
\begin{small}
\begin{align}
\textrm{R}_k(\mathbf{W},\mathbf{u}_k,\{q_k\},\bm{\phi}) = \textrm{log}(1+\textrm{SINR}_{U,k}), \forall k \in \mathcal{K}.
\end{align}
\end{small}
\end{shrinkeq}

\subsection{Problem Formulation}
In this paper,
we aim to jointly  optimize
the transmit beamformer $\mathbf{W}$,
the linear post-processing filters $\{\mathbf{u}_k\}_{0}^{K}$,
the users' UL transmission power $\{q_k\}$
and the reflection phase shift  $\bm{\phi}$
to maximize the sum-rate of all UL mobile users
while assuring the target sensing SINR remain above a predefined level.
Therefore,
the optimization problem is formulated as
\begin{shrinkeq}{-0.2em}
\begin{small}
\begin{subequations}
	\begin{align}
\textrm{(P0)}:\mathop{\textrm{max}}
\limits_{\mathbf{W},
\mathbf{u}_0,
\{\mathbf{u}_k\},
\{q_k\},
\bm{\phi}}\
&{\sum}_{k=1}^{K} \textrm{R}_k(\mathbf{W},\mathbf{u}_k,\{q_k\},\bm{\phi})\label{P0_obj}\\
\textrm{s.t.}\
&\textrm{SINR}_{r}\geq \Gamma_r,\\
&\Vert\mathbf{W}\Vert_F^2\leq P_{BS},\\
&q_k\leq P_{U,k}, \forall k \in \mathcal{K},\\
&\vert\phi_m\vert=1, \forall m \in \mathcal{M},
\end{align}
\end{subequations}
\end{small}
\end{shrinkeq}
\vspace{-0.4cm}

\noindent
where $\Gamma_r$,
$P_{BS}$
and
$P_{U,k}$
are
denoted as
the predefined level of target sensing quality,
the  transmission power budget of the BS
and
the transmission power budget of $k$-th user,
respectively.
It is obvious that the problem (P0) is highly challenging  due to its  non-convex objective and constraints.

\section{Proposed Algorithm}
\subsection{Problem Reformulation}
In order to make the problem (P0) more tractable,
we firstly employ the weighted mean square error
(WMMSE) method \cite{ref161} to equivalently  transform its objective function.
Specifically,
via introducing auxiliary variables $\{\beta_k\}$ and $\{\omega_k\}$,
the original objective function (\ref{P0_obj}) can be equivalently written into a variation form \cite{ref161},
as shown in (\ref{MSE}) on the top of this page.
\begin{figure*}
\begin{small}
\begin{align}
&\textrm{R}_k(\mathbf{W},\mathbf{u}_k,\{q_k\},\bm{\phi})\label{MSE}\\
&=\mathop{\textrm{max}}
\limits_{
\omega_k\geq0
}
\textrm{log}(\omega_k)-\omega_k\big({\sum}_{i=1}^{K}q_i\vert\mathbf{u}^H_k\mathbf{h}_{U,k}\vert^2
+\Vert\mathbf{u}^H_k\mathbf{H}\mathbf{W}\Vert^2_2
+\Vert\mathbf{u}^H_k\mathbf{G}\mathbf{W}\Vert^2_2
+\sigma_r^2\Vert\mathbf{u}^H_k\Vert^2_2\big)^{-1}\sqrt{q_k}\mathbf{u}^H_k\mathbf{h}_{U,k}+1\nonumber\\
&=\mathop{\textrm{max}}
\limits_{
\omega_k\geq0,
\beta_k
}
\underbrace{\textrm{log}(\omega_k)
-\omega_k\bigg(1\!\!-\!\!2\textrm{Re}\{\beta_k^{\ast}\sqrt{q_k}\mathbf{u}^H_k\mathbf{h}_{U,k}\}
\!\!+\!\vert\beta_k\vert^2\big({\sum}_{i=1}^{K}q_i\vert\mathbf{u}^H_k\mathbf{h}_{U,k}\vert^2
\!\!+\!\Vert\mathbf{u}^H_k\mathbf{H}\mathbf{W}\Vert^2_2
\!\!+\!\Vert\mathbf{u}^H_k\mathbf{G}\mathbf{W}\Vert^2_2
\!\!+\!\sigma_r^2\Vert\mathbf{u}^H_k\Vert^2_2\big)\bigg)+1}\limits_{\mathrm{\tilde{R}}_k(\mathbf{W},\mathbf{u}_k,\{q_k\},\bm{\phi},\omega_k,\beta_k)}, \forall k \in \mathcal{K}.\nonumber
\end{align}
\end{small}
\boldsymbol{\hrule}
\end{figure*}
Therefore,
the original problem (P0) can be reformulated as
\begin{shrinkeq}{-0.7em}
\begin{small}
\begin{subequations}
\begin{align}
\textrm{(P1)}:\mathop{\textrm{max}}
\limits_{\mathbf{W},
\mathbf{u}_0,
\{\mathbf{u}_k\},
\atop
\{\!q_k\!\},
\bm{\phi},
\{\!\omega_k\!\},
\{\!\beta_k\!\}}\
&{\sum}_{k=1}^{K}\!\! \mathrm{\tilde{R}}_k(\mathbf{W},\mathbf{u}_k,\{\!q_k\!\},\bm{\phi},\omega_k,\beta_k)\label{P1_obj}\\
\textrm{s.t.}\
&\textrm{SINR}_{r}\geq \Gamma_r,\label{P1_SINR_radar}\\
&\Vert\mathbf{W}\Vert_F^2\leq P_{BS},\label{P1_power_BS}\\
&q_k\leq P_{U,k}, \forall k \in \mathcal{K},\label{P1_power_user}\\
&\vert\phi_m\vert=1, \forall m \in \mathcal{M}\label{P1_phi_module}.
\end{align}
\end{subequations}
\end{small}
\end{shrinkeq}

In the following,
we propose to utilize the block descent ascent (BCA) \cite{ref16_BCA} method  to tackle the problem (P1).

\subsection{Optimizing Auxiliary Variables $\{\beta_k\}$, $\{\omega_k\}$}
According to the derivation of WMMSE transformation,
with given other variables,
the auxiliary variables $\{\beta_k\}$ and $\{\omega_k\}$ can be updated by the closed form solutions that are given as follows
\begin{shrinkeq}{-0.3em}
\begin{small}
\begin{align}
&\beta_k^{\star}\!\!=\!\!\frac{\sqrt{q_k}\mathbf{u}_k^H\mathbf{h}_{U,k}}
{{\sum}_{i=1}^{K}q_i\vert\mathbf{u}^H_k\mathbf{h}_{U,i}\vert^2
\!\!+\!\!\Vert\!\mathbf{u}^H_k\mathbf{H}\mathbf{W}\!\Vert^2_2
\!\!+\!\!\Vert\!\mathbf{u}^H_k\mathbf{G}\mathbf{W}\!\Vert^2_2
\!\!+\!\!\sigma_r^2\Vert\!\mathbf{u}^H_k\!\Vert^2_2},\label{Solution_beta}\\
&\omega_k^{\star}
\!\!=\!\!1\!\!+\!\!\frac{{q_k}\mathbf{h}_{U,k}^H\mathbf{u}_k\mathbf{u}_k^H\mathbf{h}_{U,k}}
{{\sum}_{i\neq k}^{K}q_i\vert\!\mathbf{u}^H_k\mathbf{h}_{U,i}\!\vert^2
\!\!+\!\!\Vert\!\mathbf{u}^H_k\!\mathbf{H}\mathbf{W}\!\Vert^2_2
\!\!+\!\!\Vert\!\mathbf{u}^H_k\!\mathbf{G}\mathbf{W}\!\Vert^2_2
\!\!+\!\!\sigma_r^2\Vert\!\mathbf{u}^H_k\!\Vert^2_2}\label{Solution_omega}.
\end{align}
\end{small}
\end{shrinkeq}

\subsection{Updating The BS Beamformer $\mathbf{W}$}
In this subsection,
we will investigate
the update of the transmit beamformer $\mathbf{W}$.
When other variables are given,
the optimization with respect to (w.r.t.)
$\mathbf{W}$ is meant to solve
\begin{shrinkeq}{-0.3em}
\begin{small}
\begin{subequations}
\begin{align}
\textrm{(P2)}:
\mathop{\textrm{min}}
\limits_{\mathbf{w}}\
&
\mathbf{w}^H\mathbf{D}_1\mathbf{w}-c_1\\
\textrm{s.t.}\
&\mathbf{w}^H\mathbf{D}_2\mathbf{w}-\mathbf{w}^H\mathbf{D}_3\mathbf{w}+c_2\leq0,\label{Beam_nonconvex_con}\\
&\mathbf{w}^H\mathbf{w}\leq P_{BS}.
\end{align}
\end{subequations}
\end{small}
\end{shrinkeq}
\vspace{-0.5cm}

\noindent
with the new parameters above defined as follows
\begin{shrinkeq}{-0.3em}
\begin{small}
\begin{align}
&\mathbf{w}\triangleq \textrm{vec}(\mathbf{W}),
c_2\triangleq
({\sum}_{k=1}^{K} q_k\vert\mathbf{u}^H_0\mathbf{h}_{U,k}\vert^2
\!+\!\sigma^2_r\Vert\mathbf{u}_0^H\Vert^2_2),\\
&\mathbf{D}_2\triangleq  \mathbf{I}_{N_t}\otimes\mathbf{G}^H\mathbf{u}_0\mathbf{u}_0^H\mathbf{G},
\mathbf{D}_3\triangleq  (\mathbf{I}_{N_t}\otimes\mathbf{H}^H\mathbf{u}_0\mathbf{u}_0^H\mathbf{H})/\Gamma_r,\nonumber\\
&c_1=
\big[{\sum}_{k=1}^{K}\big(\mathrm{log}(\omega_k)
-\omega_k
+2\mathrm{Re}\{\omega_k\beta_k^{\ast}\sqrt{q_k}\mathbf{u}^H_k\mathbf{h}_{U,k}\}\nonumber\\
&\quad-\omega_k\vert\beta_k\vert^2({\sum}_{i=1}^{K}q_i\vert\mathbf{u}^H_k\mathbf{h}_{U,i}\vert^2
+\sigma_r^2\Vert\mathbf{u}^H_k\Vert^2_2)+1\big)\big],\nonumber\\
&\mathbf{D}_1\!\triangleq\! {\sum}_{k=1}^{K}\omega_k\vert\beta_k\vert^2
(\mathbf{I}_{N_t}\!\!\otimes\!\mathbf{H}^H\mathbf{u}_k\mathbf{u}_k^H\mathbf{H}
\!+\!
\mathbf{I}_{N_t}\!\!\otimes\!\mathbf{G}^H\mathbf{u}_k\mathbf{u}_k^H\mathbf{G}).\nonumber
\end{align}
\end{small}
\end{shrinkeq}

Note that the problem (P2) is difficult
due to the difference of convex (DC) form constraint (\ref{Beam_nonconvex_con}).
Therefore,
we linearize
the concave quadratic term $\mathbf{w}^H\mathbf{D}_3\mathbf{w}$ in (\ref{Beam_nonconvex_con})
at a point ${\mathbf{w}_0}$ via the MM framework \cite{ref15} as follows
\begin{shrinkeq}{-0.3em}
\begin{small}
\begin{align}
\mathbf{w}^H\mathbf{D}_3\mathbf{w}
\geq
2\textrm{Re}\{{\mathbf{w}}^H_0\mathbf{D}_3(\mathbf{w}-{\mathbf{w}_0})\}+{\mathbf{w}}^H_0\mathbf{D}_3{\mathbf{w}}_0,\label{Beam_nonconvex_con_MM}
\end{align}
\end{small}
\end{shrinkeq}
\vspace{-0.5cm}

\noindent
where
${\mathbf{w}_0}$ is the value of ${\mathbf{w}}$ obtained in the last iteration.
Next,
we replace the  constraint (\ref{Beam_nonconvex_con}) by (\ref{Beam_nonconvex_con_MM})
and  the optimization problem (P2) is rewritten as
\begin{shrinkeq}{-0.2em}
\begin{subequations}
\begin{align}
\textrm{(P3)}:\mathop{\textrm{min}}
\limits_{\mathbf{w}}\
&
\mathbf{w}^H\mathbf{D}_1\mathbf{w}-c_1\\
\textrm{s.t.}\
&\mathbf{w}^H\mathbf{D}_2\mathbf{w}-2\mathrm{Re}\{\mathbf{d}_3^H\mathbf{w}\}+\hat{c}_2\leq0,\\
&\mathbf{w}^H\mathbf{w}\leq P_{BS},
\end{align}
\end{subequations}
\end{shrinkeq}
where
$\mathbf{d}_3 \triangleq \mathbf{D}_3^H{\mathbf{w}}_0$
and
$\hat{c}_2\triangleq{c}_2+({\mathbf{w}}^H_0\mathbf{D}_3{\mathbf{w}}_0)^{\ast}$.
The problem (P3)
is a typical second order cone program (SOCP) and can
be solved by off-the-shelf numerical solvers,
e.g.,
CVX \cite{ref17}.

\subsection{Optimizing The User Transmission Power}
With other variables being fixed,
the optimization of  all the users' transmission power $\{q_k\}$ is reduced to solve
\begin{shrinkeq}{-0.65em}
\begin{small}
\begin{subequations}
\begin{align}
\textrm{(P4)}:\mathop{\textrm{min}}
\limits_{
\{q_k\}}\
&   {\sum}_{k=1}^{K}a_k q_k + {\sum}_{k=1}^{K}b_k \sqrt{q_k}-c_3\\
\textrm{s.t.}\
& {\sum}_{k=1}^{K}d_k q_k \leq \hat{c}_3,\\
& 0 \leq q_k\leq P_{U,k}, \forall k \in \mathcal{K},
\end{align}
\end{subequations}
\end{small}
\end{shrinkeq}
\vspace{-0.3cm}

\noindent
where the newly introduced coefficients are defined as follows
\begin{shrinkeq}{-0.4em}
\begin{small}
\begin{align}
&a_k \triangleq {\sum}_{j=1}^{K}\omega_j\vert\beta_j\vert^2\vert\mathbf{u}_j^H\mathbf{h}_{U,k}\vert^2,
d_k \triangleq \vert\mathbf{u}^H_0\mathbf{h}_{U,k}\vert^2,\\
&\hat{c}_3 \triangleq  \Vert\mathbf{u}^H_0\mathbf{H}\mathbf{W}\Vert^2_2/\Gamma_r-\Vert\mathbf{u}^H_0\mathbf{G}\mathbf{W}\Vert^2_2
-\sigma^2_r\Vert\mathbf{u}_0^H\Vert^2_2,\nonumber\\
&b_k \triangleq -2\mathrm{Re}(\omega_k\beta_k^{\ast}\mathbf{u}_k^H\mathbf{h}_{U,k}),
c_3
\triangleq
{\sum}_{k=1}^{K}
\big\{\mathrm{log}(\omega_k)
-\omega_k\nonumber\\
&-\omega_k\vert\beta_k\vert^2(\Vert\mathbf{u}_k^H\mathbf{H}\mathbf{W}\Vert^2_2
+\Vert\mathbf{u}_k^H\mathbf{G}\mathbf{W}\Vert^2_2
+\sigma_r^2\Vert\mathbf{u}_k^H\Vert)+1
\big\}.\nonumber
\end{align}
\end{small}
\end{shrinkeq}

The problem (P4) can still be formulated into an SOCP problem and can be numerically solved.

\subsection{Optimizing The Users' Filters $\{\mathbf{u}_k\}_{k=1}^{K}$}
The update of $\{\mathbf{u}_k\}$ are meant to solve the following problem
\begin{shrinkeq}{-0.4em}
\begin{small}
\begin{align}
\textrm{(P5)}:\mathop{\textrm{min}}
\limits_{\{\mathbf{u}_k\}}\
& {\sum}_{k=1}^{K}(\mathbf{u}_k^H\mathbf{F}_{k}\mathbf{u}_k
\!-\!
2\textrm{Re}\{\mathbf{u}^H_k\mathbf{\tilde{h}}_{U,k}\})
\!-\!c_4
\end{align}
\end{small}
\end{shrinkeq}
\vspace{-0.3cm}

\noindent
where the parameters of (P5) are defined as
\begin{shrinkeq}{-0.4em}
\begin{small}
\begin{align}
&\mathbf{\tilde{h}}_{U,k}\! \triangleq \!\omega_k\beta_k^{\ast}\sqrt{q_k}\mathbf{h}_{U,k},
c_4
\!\triangleq\!
{\sum}_{k=1}^{K}\big(
\mathrm{log}(\omega_k)+\omega_k+1\big),\\
&\mathbf{F}_{k}
\!\!\triangleq\!\!
\omega_k\vert\beta_k\vert^2\!
\big(\!{\sum}_{i=1}^{K}\!q_i\mathbf{h}_{U,i}\mathbf{h}_{U,i}^H
\!\!+\!\!\mathbf{H}\mathbf{W}\mathbf{W}^H\!\mathbf{H}^H
\!\!\!\!+\!\!\mathbf{G}\mathbf{W}\mathbf{W}^H\!\!\mathbf{G}^H
\!\!\!+\!\!\sigma_r^2\mathbf{I}_{N_r}\!\!\big).\nonumber
\end{align}
\end{small}
\end{shrinkeq}

Hence the problem (P5) can be decomposed into $K$ independent sub-problems
with each being  given as
\begin{shrinkeq}{-0.4em}
\begin{small}
\begin{align}
\textrm{(P}\mathrm{\textrm{6}_k}\textrm{)}:\mathop{\textrm{min}}
\limits_{\mathbf{u}_k}\
& \mathbf{u}_k^H\mathbf{F}_{k}\mathbf{u}_k
-
2\textrm{Re}\{\mathbf{u}^H_k\mathbf{\tilde{h}}_{U,k}\}
\end{align}
\end{small}
\end{shrinkeq}

Notice that the problem (P$\mathrm{\textrm{6}_k}$) is a typical unconstrained convex quadratic problem.
Its optimal solution can be easily obtained via setting its derivative as zero and obtained as follows
\begin{shrinkeq}{-0.4em}
\begin{align}
\mathbf{u}_k^{\star}=\mathbf{F}_{k}^{-1}\mathbf{\tilde{h}}_{U,k}, \forall k \in \mathcal{K}.\label{Solution_u_k}
\end{align}
\end{shrinkeq}

\subsection{Optimizing Radar Filter $\mathbf{u}_0$}
When other variables are given,
the optimization problem w.r.t. $\mathbf{u}_0$ is reduced to  a feasibility characterization problem,
which is given as
\begin{shrinkeq}{-0.7em}
\begin{small}
\begin{subequations}
\begin{align}
\textrm{(P7)}:{\textrm{Find}}\
& \mathbf{u}_0\\
\textrm{s.t.}\ &
\mathbf{u}_0^H\mathbf{E}_1\mathbf{u}_0-\mathbf{u}_0^H\mathbf{E}_2\mathbf{u}_0
\leq
0.\label{Radar_SINR_c}
\end{align}
\end{subequations}
\end{small}
\end{shrinkeq}
\vspace{-0.3cm}

\noindent
where  the  parameters of problem (P7) are defined as
\begin{shrinkeq}{-0.5em}
\begin{small}
\begin{align}
&\mathbf{E}_2
\triangleq
\mathbf{H}\mathbf{W}\mathbf{W}^H\mathbf{H}^H/\Gamma_r,\\
&\mathbf{E}_1
\triangleq
\big({\sum}_{k=1}^{K}q_k\mathbf{h}_{U,k}\mathbf{h}_{U,k}^H+\mathbf{G}\mathbf{W}\mathbf{W}^H\mathbf{G}^H+\sigma_r^2\mathbf{I}_{N_r}\big).\nonumber
\end{align}
\end{small}
\end{shrinkeq}

To tackle (P7),
we consider another closely related problem as follows
\begin{shrinkeq}{-0.80em}
\begin{small}
\begin{subequations}
\begin{align}
\textrm{(P8)}:\mathop{\textrm{min}}
\limits_{\mathbf{u}_0, \alpha_{u}}\
& \alpha_{u}\\
\textrm{s.t.}\
&\mathbf{u}_0^H\mathbf{E}_1\mathbf{u}_0-\mathbf{u}_0^H\mathbf{E}_2\mathbf{u}_0
\leq
\alpha_{u}\label{Radar_SINR_c_MM}.
\end{align}
\end{subequations}
\end{small}
\end{shrinkeq}
\vspace{-0.3cm}

\noindent
Suppose that the optimal solution
($\mathbf{u}_0^{\star}$,$\alpha_u^\star$)
of (P8)
yields
$\alpha_u^{\star}\leq0$,
then
$\mathbf{u}_0^{\star}$
is actually a feasible solution to (P7).
Therefore,
we turn to solve (P8).
Note that (P8) is assured to have a solution yielding non-positive objective as long as we start iteration from a feasible solution.
It is obvious that
the operation of minimizing the variable $\alpha_{u}$ is equivalent to
minimize the left side of the constraint (\ref{Radar_SINR_c_MM}) w.r.t.  variable $\mathbf{u}_0$.
Therefore,
we turn to optimize $\mathbf{u}_0$ and  solve the following problem
\begin{shrinkeq}{-0.5em}
\begin{small}
\begin{align}
\textrm{(P9)}:\mathop{\textrm{min}}
\limits_{\mathbf{u}_0}\
& \mathbf{u}_0^H\mathbf{E}_1\mathbf{u}_0-\mathbf{u}_0^H\mathbf{E}_2\mathbf{u}_0\label{(U_0_obj)}
\end{align}
\end{small}
\end{shrinkeq}

Since the objective function (\ref{(U_0_obj)}) is non-convex due to its DC form,
we  adopt the MM method to construct a linear low-bound of $\mathbf{u}_0^H\mathbf{E}_2\mathbf{u}_0$,
which is given as
\begin{shrinkeq}{-0.6em}
\begin{small}
\begin{align}
\mathbf{u}_0^H\mathbf{E}_2\mathbf{u}_0
&\geq
\hat{\mathbf{u}}_0^H\mathbf{E}_2\hat{\mathbf{u}}_0
+2\textrm{Re}\{\hat{\mathbf{u}}_0^H\mathbf{E}_2(\mathbf{u}_0-\hat{\mathbf{u}}_0)\}\nonumber\\
&=
2\textrm{Re}\{\hat{\mathbf{u}}_0^H\mathbf{E}_2\mathbf{u}_0\}
-
(\hat{\mathbf{u}}_0^H\mathbf{E}_2\hat{\mathbf{u}}_0)^{\ast},\label{U_0_linear}
\end{align}
\end{small}
\end{shrinkeq}
\vspace{-0.4cm}

\noindent
where $\hat{\mathbf{u}}_0$ is feasible solution obtained in the last iteration.

Then replacing the term $\mathbf{u}_0^H\mathbf{E}_2\mathbf{u}_0$  by  (\ref{U_0_linear}),
we turn to optimize a tight convex upper bound of the objective (P9),
which is given as
\begin{shrinkeq}{-0.5em}
\begin{align}
\textrm{(P10)}:\mathop{\textrm{min}}
\limits_{\mathbf{u}_0}\
& \mathbf{u}_0^H\mathbf{E}_1\mathbf{u}_0\!-\!2\textrm{Re}\{\hat{\mathbf{u}}_0^H\mathbf{E}_2\mathbf{u}_0\}
\!+\!
(\hat{\mathbf{u}}_0^H\mathbf{E}_2\hat{\mathbf{u}}_0)^{\ast}
\end{align}
\end{shrinkeq}
The problem (P10) is a unconstraint convex quadratic problem
and its optimal solution is directly given  as
\begin{shrinkeq}{-0.6em}
\begin{align}
\mathbf{u}_0^{\star}
=
\mathbf{E}_1^{-1}(\mathbf{E}_2^H\hat{\mathbf{u}}_0).\label{Solution_u_0}
\end{align}
\end{shrinkeq}

\subsection{Optimizing The Phase Shift $\bm{\phi}$ }
In this subsection,
we investigate the optimization of
the phase shift $\bm{\phi}$.
By introducing the new definitions as follows
\begin{shrinkeq}{-0.3em}
\begin{small}
\begin{align}
&\mathbf{P}_k
\triangleq
\mathbf{G}_r^H\mathrm{diag}(\mathbf{h}_{RU,k}),
\mathbf{r}_k
\triangleq
\mathbf{G}_r\mathbf{u}_k,\\
&\mathbf{S}_k
\triangleq
\mathbf{W}^H\mathbf{G}_t^H\mathrm{diag}(\mathbf{r}_k),
\mathbf{v}_k
\triangleq
\mathbf{u}_k^H\mathbf{H}_{s}^H\mathbf{W},\nonumber\\
&\mathbf{r}_0
\triangleq
\mathbf{G}_r\mathbf{u}_0,
\mathbf{S}_0
\triangleq
\mathbf{W}^H\mathbf{G}_t^H\mathrm{diag}(\mathbf{r}_0),
\mathbf{v}_0
\triangleq
\mathbf{u}_0^H\mathbf{H}_{s}^H\mathbf{W},\nonumber
\end{align}
\end{small}
\end{shrinkeq}
\vspace{-0.5cm}

\noindent
the objective function (\ref{P1_obj}) and the constraint (\ref{P1_SINR_radar}) can be respectively rewritten as
\begin{shrinkeq}{-0.2em}
\begin{subequations}
\begin{align}
&-{\sum}_{k=1}^{K} \mathrm{\tilde{R}}_k
= \bm{\phi}^H\mathbf{T}\bm{\phi}
-2\mathrm{Re}\{\mathbf{x}^H\bm{\phi}\}-c_5,\label{Phi_obj_trans}\\
&\bm{\phi}^H\mathbf{T}_0\bm{\phi}
-2\textrm{Re}\{\mathbf{x}_0^H\bm{\phi}\}+c_6\leq 0,\label{Phi_SINR_trans}
\end{align}
\end{subequations}
\end{shrinkeq}
with the parameters in (\ref{Phi_obj_trans}) and (\ref{Phi_SINR_trans})
being defined in (\ref{Phi_trans_coefficient}),
as shown on the top of next page.
\begin{figure*}
\begin{small}
\begin{align}
&c_6
\triangleq
({\sum}_{k=1}^{K}q_k\mathbf{u}_0^H\mathbf{h}_{BU,k}\mathbf{h}_{BU,k}^H\mathbf{u}_0
+
\mathbf{v}_0\mathbf{v}^H_0+\sigma_r^2\Vert\mathbf{u}_0^H\Vert^2_2)-\Vert\mathbf{u}_0^H\mathbf{H}\mathbf{W}\Vert_2^2/{\Gamma_r},\label{Phi_trans_coefficient}\\
&\mathbf{T}
\triangleq
{\sum}_{k=1}^{K}
\big\{
\omega_k\vert\beta_k\vert^2
\big[
{\sum}_{i=1}^{K}q_i(\mathbf{P}_{i}^H\mathbf{u}_k\mathbf{u}_k^H\mathbf{P}_{i})
+
\mathbf{S}_k^T\mathbf{S}_k^{\ast}
\big]
\big\},
\mathbf{T}_0
\triangleq
\big({\sum}_{k=1}^{K}q_k\mathbf{P}_{k}^H\mathbf{u}_0\mathbf{u}_0k^H\mathbf{P}_{k}
+
\mathbf{S}_0^T\mathbf{S}_0^{\ast}\big),\nonumber
\\
&\mathbf{x}_0
\triangleq
-\big({\sum}_{k=1}^{K}q_k
(
\mathbf{h}_{BU,k}^H\mathbf{u}_0\mathbf{u}_0^H\mathbf{P}_{k})
+
\mathbf{v}_0^{\ast}\mathbf{S}_0^{\ast}\big)^H,
\mathbf{x}
\triangleq
{\sum}_{k=1}^{K}
\big\{
\omega_k\beta_k^{\ast}\sqrt{q_k}\mathbf{u}_k^H\mathbf{P}_k
-
\omega_k\vert\beta_k\vert^2
\big[
{\sum}_{i=1}^{K}q_i
(
\mathbf{h}_{BU,i}^H\mathbf{u}_k\mathbf{u}_k^H\mathbf{P}_{i})
+
\mathbf{v}_k^{\ast}\mathbf{S}_k^{\ast}
\big]
\big\}^H,\nonumber\\
&c_5
\triangleq
{\sum}_{k=1}^{K}
\{
\mathrm{log}(\omega_k)-\omega_k+1
\!+\!
2\mathrm{Re}\{\omega_k\beta_k^{\ast}\sqrt{q_k}\mathbf{u}_k^H\mathbf{h}_{BU,k}\}
-\omega_k\vert\beta_k\vert^2
[
{\sum}_{i=1}^{K}q_i\mathbf{u}_k^H\mathbf{h}_{BU,i}\mathbf{h}_{BU,i}^H\mathbf{u}_k
\!+\!
\Vert\mathbf{u}_k^H\mathbf{H}\mathbf{W}\Vert^2_2
\!+\!
\mathbf{v}_k\mathbf{v}_k^H
\!+\!
\sigma_r^2\Vert\mathbf{u}_k^H\Vert^2_2
]
\}\nonumber.
\end{align}
\end{small}
\boldsymbol{\hrule}
\end{figure*}

Based on the above transformation,
the optimization problem w.r.t. $\bm{\phi}$ is reduced to
\begin{shrinkeq}{-0.7em}
\begin{small}
\begin{subequations}
\begin{align}
\textrm{(P11)}:\mathop{\textrm{min}}
\limits_{
\bm{\phi}}\
& \bm{\phi}^H\mathbf{T}\bm{\phi}
-2\textrm{Re}\{\mathbf{x}^H\bm{\phi}\}
-c_5\\
\textrm{s.t.}\
&\bm{\phi}^H\mathbf{T}_0\bm{\phi}
-2\textrm{Re}\{\mathbf{x}_0^H\bm{\phi}\}
+c_6\leq 0,\\
&\vert\phi_m\vert=1, \forall m \in \mathcal{M}\label{Phi_c_module_c}.
\end{align}
\end{subequations}
\end{small}
\end{shrinkeq}
\vspace{-0.2cm}

Due to the  nonlinear equality constraint (\ref{Phi_c_module_c}),
the problem (P11) is non-convex.
To resolve (P11),
we adopt the PDD method  \cite{ref16}.
Firstly,
to decouple the nonconvex constraints (\ref{Phi_c_module_c}),
we introduce an auxiliary variable $\bm{\psi}$ and transform problem (P11) as follows
\begin{shrinkeq}{-0.6em}
\begin{small}
\begin{subequations}
\begin{align}
\textrm{(P12)}:\mathop{\textrm{min}}
\limits_{
\bm{\phi},
\bm{\psi}
}\
& \bm{\phi}^H\mathbf{T}\bm{\phi}
-2\mathrm{Re}\{\mathbf{x}^H\bm{\phi}\}
-c_5\\
\textrm{s.t.}\
&\bm{\phi}^H\mathbf{T}_0\bm{\phi}
-2\textrm{Re}\{\mathbf{x}_0^H\bm{\phi}\}
+c_6\leq 0,\\
&\bm{\phi}=\bm{\psi},\label{P11_equlity_constraint}\\
&\vert\psi_m\vert=1, \forall m \in \mathcal{M}.
\end{align}
\end{subequations}
\end{small}
\end{shrinkeq}
\vspace{-0.2cm}

Next,
via penalizing the equality constraint (\ref{P11_equlity_constraint}),
the augmented Lagrangian (AL) problem of (P12) is given as
\begin{shrinkeq}{-0.7em}
\begin{small}
\begin{subequations}
\begin{align}
\textrm{(P13)}:\mathop{\textrm{min}}
\limits_{
\bm{\phi},
\bm{\psi},
\rho,
\bm{\lambda}
}\
& \bm{\phi}^H\mathbf{T}\bm{\phi}
-2\textrm{Re}\{\mathbf{x}^H\bm{\phi}\}
-c_5\\
&+
\frac{1}{2\rho}\Vert\bm{\phi}-\bm{\psi}\Vert_2^2
+
\textrm{Re}\{\bm{\lambda}^H(\bm{\phi}-\bm{\psi})\}\nonumber
\\
\mathrm{s.t.}\
&\bm{\phi}^H\mathbf{T}_0\bm{\phi}
-2\textrm{Re}\{\mathbf{x}_0^H\bm{\phi}\}
+c_6\leq 0,\\
&\vert\psi_m\vert=1, \forall m \in \mathcal{M}.
\end{align}
\end{subequations}
\end{small}
\end{shrinkeq}
\vspace{-0.4cm}

\noindent
where $\rho$ is the penalty coefficient
and $\bm{\lambda}$ is the Lagrangian multiplier associated with (\ref{P11_equlity_constraint}).
Guided by the PDD framework \cite{ref16},
we conduct a two-layer iteration procedure,
with its inner layer alternatively updating $\bm{\phi}$ and $\bm{\psi}$
in a  block coordinate descent (BCD) manner
while the outer layer selectively updating the penalty coefficient $\rho$ or the dual variable $\bm{\lambda}$.
The PDD  procedure will be elaborated in the following.

\normalem
\underline{\emph{Inner Layer Procedure}}

For the inner layer iteration,
we alternatively update $\bm{\phi}$ and $\bm{\psi}$.
When $\bm{\psi}$ is given,
the minimization of AL  w.r.t.   $\bm{\phi}$ reduces to solving the following problem
\begin{shrinkeq}{-0.6em}
\begin{small}
\begin{subequations}
\begin{align}
\textrm{(P14)}:\mathop{\textrm{min}}
\limits_{
\bm{\phi}
}\
& \bm{\phi}^H\mathbf{T}\bm{\phi}
-2\textrm{Re}\{\mathbf{x}^H\bm{\phi}\}-c_5\\
&
+
\frac{1}{2\rho}\Vert\bm{\phi}-\bm{\psi}\Vert_2^2
+
\textrm{Re}\{\bm{\lambda}^H(\bm{\phi}-\bm{\psi})\}\nonumber
\\
\textrm{s.t.}\
&\bm{\phi}^H\mathbf{T}_0\bm{\phi}
-2\textrm{Re}\{\mathbf{x}_0^H\bm{\phi}\}
+c_6\leq 0.
\end{align}
\end{subequations}
\end{small}
\end{shrinkeq}
\vspace{-0.5cm}

\noindent
The problem (P14) is an SOCP and can be solved by CVX.

When $\bm{\phi}$ is fixed,
the update of the auxiliary variable $\bm{\psi}$ is meant to solve
\begin{shrinkeq}{-0.7em}
\begin{small}
\begin{subequations}
\begin{align}
\textrm{(P15)}:\mathop{\textrm{min}}
\limits_{
\bm{\psi}
}\
&
\frac{1}{2\rho}\Vert\bm{\phi}-\bm{\psi}\Vert_2^2
+
\textrm{Re}\{\bm{\lambda}^H(\bm{\phi}-\bm{\psi})\}\label{Psi_obj}
\\
\textrm{s.t.}\
&\vert\psi_m\vert=1, \forall m \in \mathcal{M}.
\end{align}
\end{subequations}
\end{small}
\end{shrinkeq}

It is readily seen that the quadratic term w.r.t. $\bm{\psi}$ in the objective function (\ref{Psi_obj}) is constant,
i.e.,
$\Vert\bm{\psi}\Vert^2_2/(2\rho)=M/(2\rho)$ since $\bm{\psi}$ has unit modulus entries.
Therefore,
the problem (P15) is equivalently written as
\begin{shrinkeq}{-0.7em}
\begin{small}
\begin{align}
\textrm{(P16)}:\mathop{\textrm{max}}
\limits_{
\vert\bm{\psi}\vert=\mathbf{1}_M
}\
&\textrm{Re}\{(\bm{\phi}+\rho\bm{\lambda})^H\bm{\psi}\}
\end{align}
\end{small}
\end{shrinkeq}

Note that the maximum of problem (P16) can be achieved
when the phases of the entries of $\bm{\psi}$
are all aligned with those of the linear coefficient $(\rho^{-1}\bm{\phi}+\bm{\lambda})$,
that the optimal solution is given as
\begin{shrinkeq}{-0.7em}
\begin{small}
\begin{align}
\bm{\psi}^{\star}=\textrm{exp}\big(j\cdot\angle(\bm{\phi}+\rho\bm{\lambda})\big)\label{PSI_Solution}.
\end{align}
\end{small}
\end{shrinkeq}

The objective value of (P13) will achieve monotonically converge
by  alternatively updating  $\bm{\phi}$ and $\bm{\psi}$ via BCD method in the inner layer.

\normalem
\underline{\emph{Outer Layer Procedure}}

In the outer layer,
we adjust the value of dual variable $\bm{\lambda}$ or the penalty coefficient $\rho$.
Specifically,
\begin{enumerate}[1)]
\item when  $\bm{\phi}=\bm{\psi}$
is approximately achieved, i.e.,
$\Vert\bm{\psi}-\bm{\phi}\Vert_{\infty} $
is smaller than some predefined diminishing threshold  $\eta_k$ \cite{ref16},
the dual variable $\bm{\lambda}$ will be updated in a gradient ascent manner as follows:
\begin{shrinkeq}{-0.5em}
\begin{small}
\begin{align}
  \bm{\lambda}^{(k+1)}:=\bm{\lambda}^{(k)}+\rho^{-1}(\bm{\phi}-\bm{\psi});
\end{align}
\end{small}
\end{shrinkeq}
\item if the equality constraint $\bm{\phi}=\bm{\psi}$ is far from ``being true",
for achieving $\bm{\phi}=\bm{\psi}$ in the subsequent iterations,
the outer layer will increase the penalty parameter $\rho^{-1}$ as follows:
\begin{shrinkeq}{-0.5em}
\begin{small}
\begin{align}
\big(\rho^{(k+1)}\big)^{-1}:=c^{-1}\cdot\big(\rho^{(k)}\big)^{-1},
\end{align}
\end{small}
\end{shrinkeq}
\vspace{-0.5cm}

\noindent
where $c$ is  a positive constant,
whose value is smaller than 1 and typically chosen in the range of [0.8, 0.9].
\end{enumerate}

The PDD-based method to solve problem (P11) is summarized in Algorithm \ref{alg:PDD}.
The overall algorithm to solve problem (P1) is  specified in Algorithm \ref{alg:Overall}.
\begin{algorithm}[t]
\caption{PDD Method to Solve (P11)}
\label{alg:PDD}
\begin{algorithmic}[1]
\STATE {initialize}
$\bm{\phi}^{(0)}$,
$\bm{\psi}^{(0)}$,
$\bm{\lambda}^{(0)}$,
${\rho}^{(0)}$
and
$k=1$
;
\REPEAT
\STATE set $\bm{\phi}^{(k-1,0)}:=\bm{\phi}^{(k-1)}$,
$\bm{\psi}^{(k-1,0)}:=\bm{\psi}^{(k-1)}$, $t=0$;
\REPEAT
\STATE  update $\bm{\phi}^{(k-1,t+1)}$  by  solving (P14);
\STATE  update $\bm{\psi}^{(k-1,t+1)}$ by (\ref{PSI_Solution});
\STATE  $t++$;
\UNTIL{$convergence$}
\STATE set $\bm{\phi}^{(k)}:=\bm{\phi}^{(k-1,\infty)}$,
$\bm{\psi}^{(k)}:=\bm{\psi}^{(k-1,\infty)}$;
\IF{$\Vert\bm{\phi}^{(k)}-\bm{\psi}^{(k)}\Vert_{\infty}\leq\eta_k $ }
\STATE{ $\bm{\lambda}^{(k+1)}:=\bm{\lambda}^{(k)}+\dfrac{1}{{\rho}^{(k)}}(\bm{\phi}^{(k)}-\bm{\psi}^{(k)})$, ${\rho}^{(k+1)}:= {\rho}^{(k)}$};
\ELSE
\STATE{$\bm{\lambda}^{(k+1)}:=\bm{\lambda}^{(k)}$, $1/{\rho}^{(k+1)}:= 1/(c\cdot{\rho}^{(k)})$};
\ENDIF \STATE $k++$;
\UNTIL{$\Vert\bm{\phi}^{(k)}-\bm{\psi}^{(k)}\Vert_{2}$
 is sufficiently small;}
\end{algorithmic}
\end{algorithm}
\begin{algorithm}[t]
\caption{Proposed Algorithm to Solve (P1)}
\label{alg:Overall}
\begin{algorithmic}[1]
\STATE {randomly generate feasible}
$\bm{\phi}^{(0)}$,
$\mathbf{W}^{(0)}$,
$\{q_k^0\}$,
$\{\mathbf{u}_k\}$,
$\mathbf{u}_0$,
and
$i=0$;
\REPEAT
\STATE update $\{\beta_k^{(i+1)}\}$ and $\{\omega_k^{(i+1)}\}$ by (\ref{Solution_beta}) and (\ref{Solution_omega}), respectively.
\STATE set $\mathbf{W}^{(i,0)} := \mathbf{W}^{(i)}$, $m=0$;
\REPEAT
\STATE update $\mathbf{W}^{(i,m+1)}$ by solving  (P3);
\STATE $m++$;
\UNTIL{$convergence$;}
\STATE set $\mathbf{W}^{(i+1)} := \mathbf{W}^{(i,\infty)}$;
\STATE update $\{{q}_k^{(i+1)}\}$ by solving (P4);
\STATE update $\{\mathbf{u}_k^{(i+1)}\}$ by (\ref{Solution_u_k});
\STATE set $\mathbf{u}_0^{(i,0)} := \mathbf{u}_0^{(i)}$, $m=0$;
\REPEAT
\STATE update $\mathbf{u}_0^{(i,m+1)}$ by (\ref{Solution_u_0});
\STATE $m++$;
\UNTIL{$convergence$;}
\STATE set $\mathbf{u}_0^{(i+1)} := \mathbf{u}_0^{(i,\infty)}$;
\STATE update $\bm{\phi}^{(i+1)}$ by invoking Alg.1;
\STATE $i++$;
\UNTIL{$convergence$;}
\end{algorithmic}
\end{algorithm}

\section{Numerical Results}
\begin{figure*}
\centering
\begin{minipage}[c]{0.3\linewidth}
\centering
\includegraphics[scale=0.12]{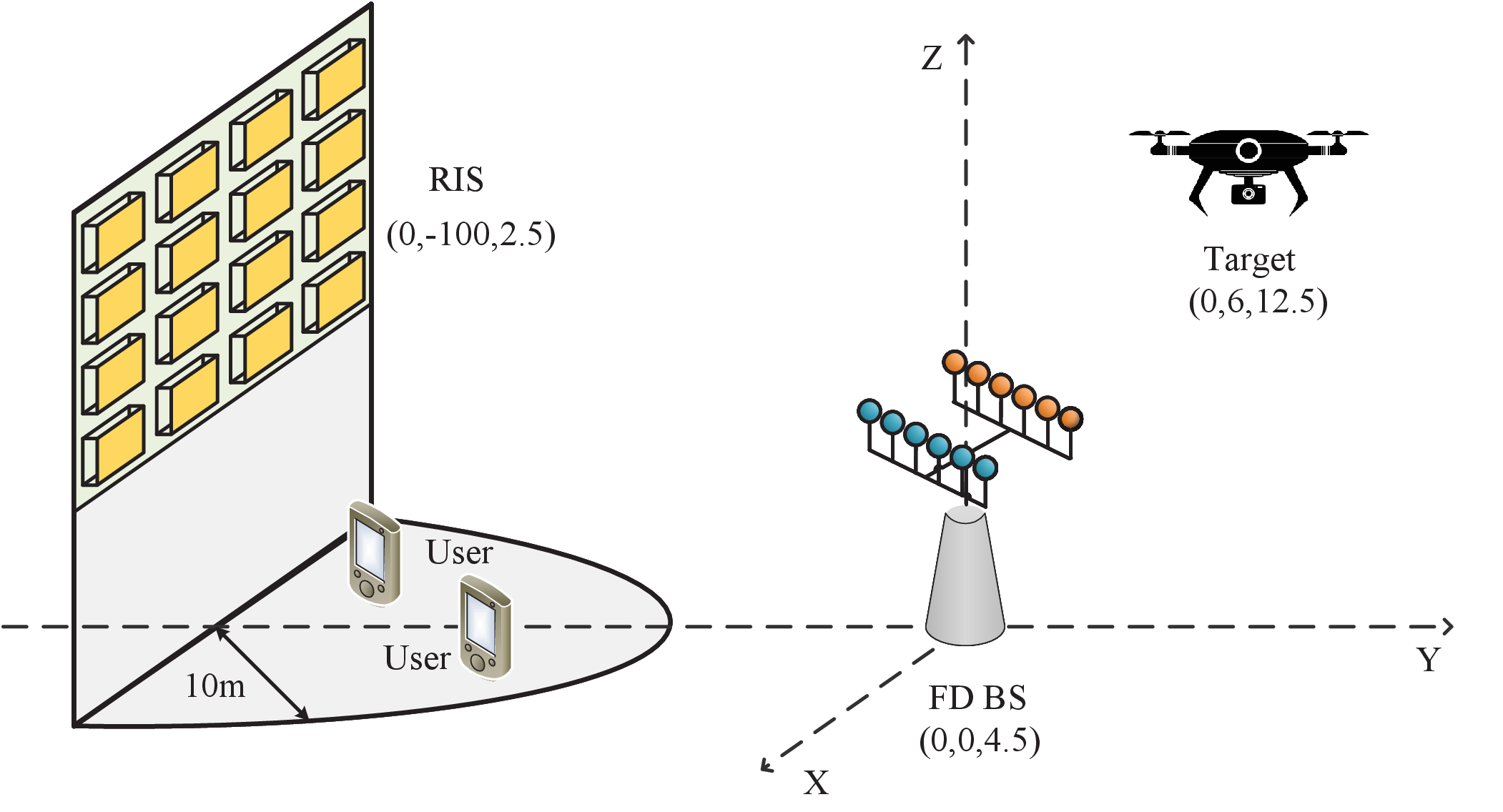}
	\caption{The experiment setting.}
	\label{fig.2}
\end{minipage}
\begin{minipage}[c]{0.3\linewidth}
\centering
\includegraphics[scale=0.28]{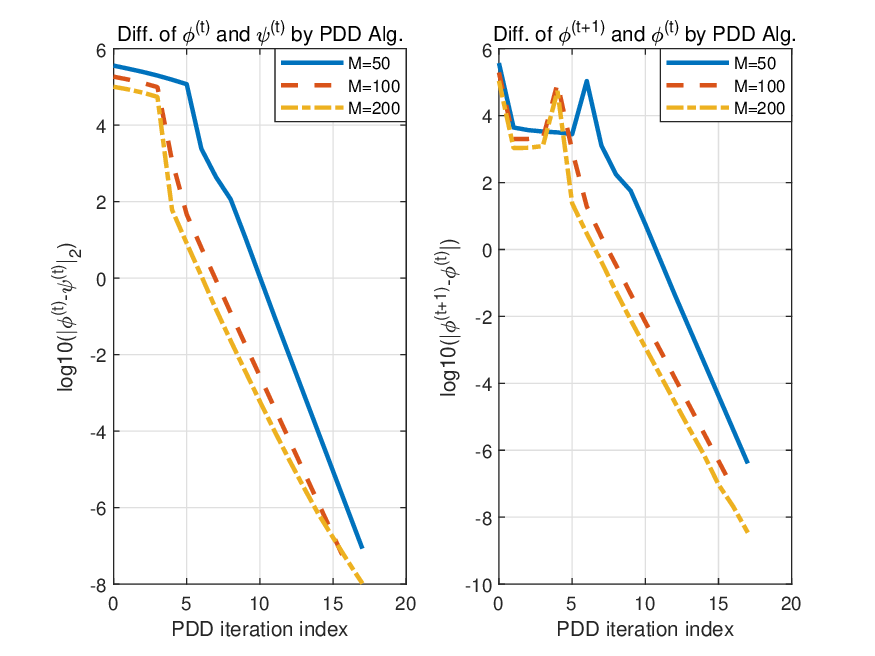}
	\caption{Convergence of  PDD method.}
	\label{fig.3}
\end{minipage}
\end{figure*}
In this section,
we will provide
numerical results to verify the performance of our proposed algorithm.
The setting of the experiment is shown in Fig. \ref{fig.2},
where one FD BS
serves $4$ users aided by an RIS and attempts to detect $1$ point target simultaneously.
In the experiment,
We assume that the locations of  BS, RIS and  target  are
($0$,$0$,$4.5$m),
($0$,$-100$m,$2.5$m)
and
($0$,$6$m,$12.5$m),
respectively.
Besides,
all users are randomly distributed within a right half circle of the radius of $10$m centered at the RIS at an altitude of $1.5$m.
The SI and BS-RIS  links are assumed to be Rician fading channel model with  Rician factor of $5$dB and $4$dB, respectively.
The BS-User links and RIS-User links all are assumed to be Rayleigh fading channels.
The BS-Target/Target-BS links are modeled
as line-of-sight (LoS) channels.
The path loss  exponents of
BS-User,
BS-RIS,
RIS-User,
BS-Target
and
Target-BS
are
$\alpha_{BU} = 3.6$,
$\alpha_{BR} = 2.7$,
$\alpha_{RU} = 2.4$
and
$\alpha_{BT} = \alpha_{TB} = 2.2$,
respectively.
The path loss of  SI channel is set to
$\rho_{SI} = -110$dB due to the SI cancellation  \cite{ref18}.
In addition,
we assume that the RIS consists of $M = 100$ units and the BS equips $N_t = N_r = 4$ transmit and receive antennas,
respectively.
Other parameters setting are $\textrm{P}_{BS}=30$dBm,
$\sigma_{BS}^2  = -90$dBm,
$\Gamma_r = 5$dB
and
$\sigma_{t}^2 = 1$.


Fig. \ref{fig.3}
illustrates the converge behaviours of our proposed  PDD-based method
for updating phase shift $\bm{\phi}$.
In Fig. \ref{fig.3},
under various settings of RIS units $M$,
the left and right sub-figure show the difference between $\bm{\phi}$ and $\bm{\psi}$
and the variation in $\bm{\phi}$ itself along with iteration progress in log domain.
As reflected by Fig. \ref{fig.3},
with the execution of the PDD method,
the variation in $\bm{\phi}$ and difference between $\bm{\phi}$ and $\bm{\psi}$
become negligible (below $10^{-6}$) within $20$ iterations.

\begin{figure*}
\centering
\begin{minipage}[c]{0.3\linewidth}
\centering
\includegraphics[scale=0.28]{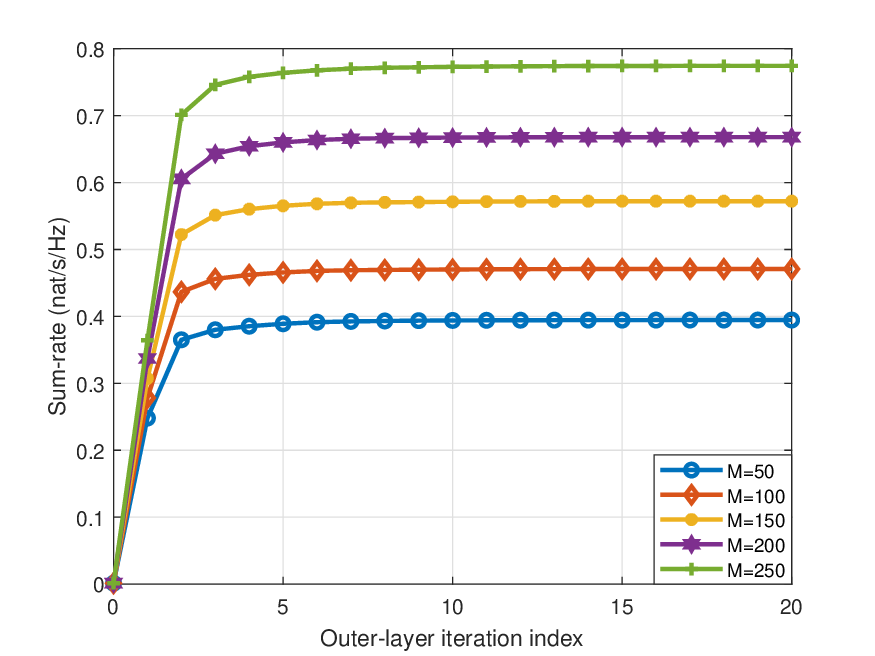}
	\caption{Convergence of Alg. 2.}
	\label{fig.4}
\end{minipage}
\begin{minipage}[c]{0.3\linewidth}
\centering
\includegraphics[scale=0.28]{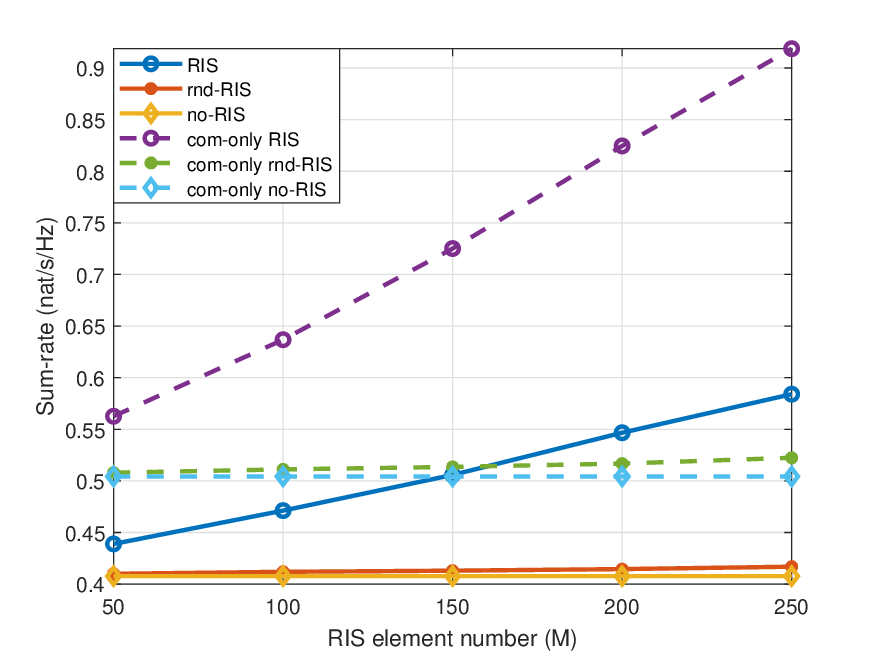}
	\caption{The impact of  $M$.}
	\label{fig.5}
\end{minipage}
\begin{minipage}[c]{0.3\linewidth}
\centering
\includegraphics[scale=0.28]{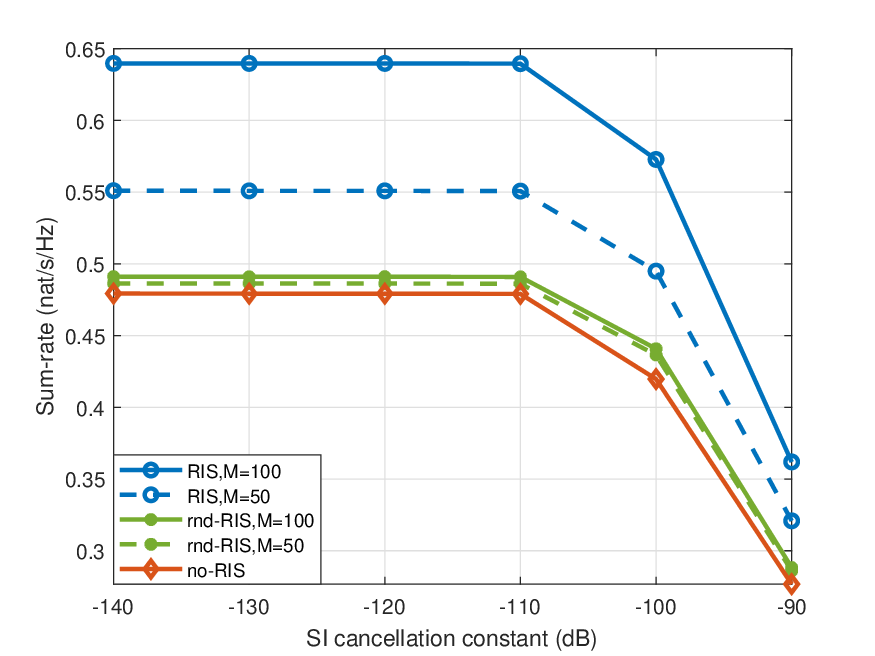}
	\caption{The impact of SI channel.}
	\label{fig.6}
\end{minipage}
\end{figure*}


Fig. \ref{fig.4}  demonstrates  the convergence behaviours of our proposed algorithm with different numbers of RIS' elements.
For each specific $M$,
the algorithm generally converges  within $10$ iterations and yield monotonic improvement in sun-rate.


In Fig. \ref{fig.5},
we present the sum-rate versus the number of RIS units.
For comparison,
we  consider the communication-only (``com-only") system.
It is clearly shown that increasing the number of RIS elements can boost the sum-rate performance.
Moreover,
our proposed algorithm (``RIS'') significantly outperforms both  no RIS (``no-RIS") and  random phase-shift RIS (``rnd-RIS") schemes in ISAC and communication-only systems,
respectively.
Besides,
due to the trade-off between the communication and radar sensing performance,
we can observe the performance gap between ISAC and communication-only system.


In Fig. \ref{fig.6},
we examine the impact of the magnitude of SI channel.
The horizontal axis shows the SI coefficient $\rho_{SI}$,
which is proportional to the magnitude of SI channel $\mathbf{H}_{S}$.
As can be seen,
sum-rate drops when SI increases.
Compared to the no-RIS scenario,
the deployment of RIS significantly boosts the sum-rate.

\section{Conclusions}
In this paper,
we investigate the joint active and passive beamforming design problem in an RIS aided FD ISAC system
that performs target sensing and UL communication functionality simultaneously.
Via utilizing MM and PDD methods,
we develop an iterative solution that efficiently updates all variables via convex optimization techniques.
Numerical results demonstrate the effectiveness of our proposed algorithm
and the benefit of deploying RIS in the FD ISAC system.


\end{document}